\documentclass[10pt,letterpaper]{article}
\usepackage[utf8]{inputenc}
\usepackage{authblk}
\usepackage[]{graphicx}
\usepackage{amsmath}
\usepackage{cite}

\title{Diagonal Slice Four-Wave Mixing: Natural Separation of Coherent Broadening Mechanisms}
\author[1 ]{Geoffrey M. Diederich}
\author[2 ]{Travis M. Autry}
\author[1,*]{Mark E. Siemens}

\affil[1 ]{Department of Physics and Astronomy, University of Denver, 2112 East Wesley Avenue, Denver, Colorado 80208, USA}
\affil[2 ]{National Institute of Standards and Technology, 325 Broadway, Boulder, Colorado 80305, USA}
\affil[*]{Corresponding author: Mark.Siemens@du.edu}

\begin{document}

\maketitle
\begin{abstract}

We present an ultrafast coherent spectroscopy data acquisition scheme that samples slices of the time domain used in multidimensional coherent spectroscopy to achieve faster data collection than full spectra. We derive analytical expressions for resonance lineshapes using this technique that completely separate homogeneous and inhomogeneous broadening contributions into separate projected lineshapes for arbitrary inhomogeneous broadening. These lineshape expressions are also valid for slices taken from full multidimensional spectra and allow direct measurement of the parameters contributing to the lineshapes in those spectra as well as our own.
\end{abstract}

Multidimensional Coherent Spectroscopy (MDCS) is a powerful spectroscopic tool for measuring dephasing and coherent dynamics of electronic and vibrational resonances on ultrafast timescales \cite{Cundiff2008,Jonas2003,Cho2008,Nardin2016,Boyd2008}. MDCS was first implemented in nuclear magnetic resonance (NMR) experiments using radio frequencies \cite{Aue1976,Jeener1979a,Ernst1987,ErnstNobel},
was then used to study vibrational coupling in the IR \cite{Hamm1998}, and has recently been extended to electronic resonances in the visible and phonons in solids at terahertz frequencies \cite{Hybl1998,Kuehn2011,Kuehn2009}. 

An advantange of MDCS is that the coupling between resonances becomes obvious and seperable because they are spread across multiple frequency dimensions; a result not available in linear spectroscopy. Because of this, MDCS provides a clear visualization of the qualitative dynamics of the system and rich quantitative information on material properties that are accessible through analysis of the resonance lineshapes present in the spectra. For example, homogeneous and inhomogeneous mechanisms broaden the resonance in orthogonal directions in a MDCS spectrum, allowing straightforward separation and identification of these contributions \cite{Bristow2011}.


While powerful and intuitive, 
it is difficult to quantify material properties due to the 
increased complexity of 
multidimensional lineshapes.  
One dimensional lineshape analysis has historically enabled measurement of the oscillator strength \cite{Belousov1995}, excitation lifetime \cite{Stanley1991,Wegener1990}, homogeneous and inhomogeneous linewidths \cite{Noll1990,Schwab1991,Erland1994}, and chemical shifts \cite{Au-Yeung1983}.
The most common approach to optical multidimensional data sets employs quasi one-dimensional lineshape analysis of frequency-domain slices \cite{Siemens2010a}, although other approaches exist \cite{Bell2015,Nagayama1978,Tokmakoff2000a}. 
That method required taking data slices from the frequency domain for lineshape fitting. The frequency slice approach can be used to measure changes in the homogeneous response of an inhomogeneous distribution \cite{Cundiff2012a,Cundiff2012,Bristow2011}.  However, all extracted slices depend on both homogeneous and inhomogeneous broadening. This problem can be mitigated by simultaneous fitting of diagonal and cross-diagonal lineshapes or fitting entire MDCS spectra \cite{Siemens2010a,Bell2015}. However, to our knowledge, no MDCS lineshape analysis has completely separated homogeneous and inhomogeneous broadening. 



In this letter, we present diagonal slice four-wave mixing (DS-FWM), 
a data acquisition scheme for rapidly measuring material properties accessible in a MDCS spectrum without acquiring a full MDCS data set.

Critically, our analysis determines both a time-domain basis and frequency-domain analytic functions that are orthogonal with respect to the broadening mechanisms found in MDCS by utilizing time slices from the rephasing pulse sequence in either the diagonal or 
cross-diagonal directions. We derive analytical expressions for the complex signal response in which different broadening mechanisms are decoupled along different time axes, using perturbative solutions to the optical Bloch equations (OBEs). Analytical expressions for frequency domain projections are derived by applying the projection-slice theorem to the time-domain slices. The resulting frequency-domain expressions completely separate the homogeneous and inhomogeneous broadening, fit simulated resonances and experimental data from \textit{GaAs} quantum wells (QWs), and demonstrate excellent agreement with previously used lineshape analysis.


The Projection-Slice Theorem states that the Fourier transform of a slice in two dimensions is equivalent to a projection onto that axis in the Fourier domain \cite{Nagayama1978}. Mathematically,
\begin{equation}\label{eq:ProjSlice}
 P(k_x) =\mathcal{F}[S(x)]=\int_{-\infty}^{\infty}S(x)e^{-i2\pi k_x x}dx,
\end{equation}
where $S(x)$ denotes a slice in any arbitrary $x$ direction and $P(k_{x})$ denotes a projection onto the same $k_{x}$ axis in the Fourier domain. 
We consider only the rephasing pulse sequence
for a sample with inhomogeneous broadening.
This system will exhibit a photon-echo at $t=\tau$, where $\tau,\, T=0, \, t$ are the inter-pulse delays between the first-second ($\tau$), second-third ($T$), and third-fourth ($t$) pulses. Note, in the present analysis the signal is assumed to be a third order coherence mapped onto a fourth order population by a fourth pulse in contrast to heterodyning schemes \cite{Martin2018,Borde1984}.

The perturbative time-domain solution to the OBEs for an inhomogeneously broadened ensemble of two-level systems interacting with this pulse sequence is;
\begin{equation} \label{eq:Signalttau}
s(t,\tau)=s_{0,0} e^{ - \left ( \gamma \left ( t + \tau  \right ) + i \omega_{0} \left ( t - \tau \right ) + \sigma^2 \left ( t - \tau \right )^2 /2 \right )} \Theta(t) \Theta(\tau),
\end{equation} 
where $s_{0,0}$ is the signal amplitude at time zero, $\omega_{0}$ is the center frequency of the resonance, $\gamma$ and $\sigma$ are the homogeneous and inhomogeneous dephasing rates, $\Theta$ denotes a unit step function that enforces causality between the pulses and the signal, and $\tau$ ($t$) is the time delay corresponding to absorption (emission) processes. In the frequency domain, $\omega_{\tau}$ ($\omega_{t}$) is the frequency axis for absorption (emission) processes. Recent work has extended the formalism of spectral analysis to include non-delta function pulses \cite{Smallwood2016}, pulses with chirp \cite{Kohler2017b}, and non-Gaussian responses \cite{Farag2017}. However, in this study these considerations are excluded in that delta function pulses and Gaussian inhomogeneous broadening are assumed. 

We can now rewrite Eq. (\ref{eq:ProjSlice}) in terms of physical parameters relevant to MDCS data,
\begin{equation} \label{eq:ProjSliceMDCS}
P(\omega_{t}+\omega_{\tau}) =\int_{-\infty}^{\infty}S(t+\tau)e^{i2\pi (t+\tau) (\omega_{t}+\omega_{\tau})}d(t+\tau).
\end{equation} 

\begin{figure}

   \centering
   \includegraphics[width=\linewidth]{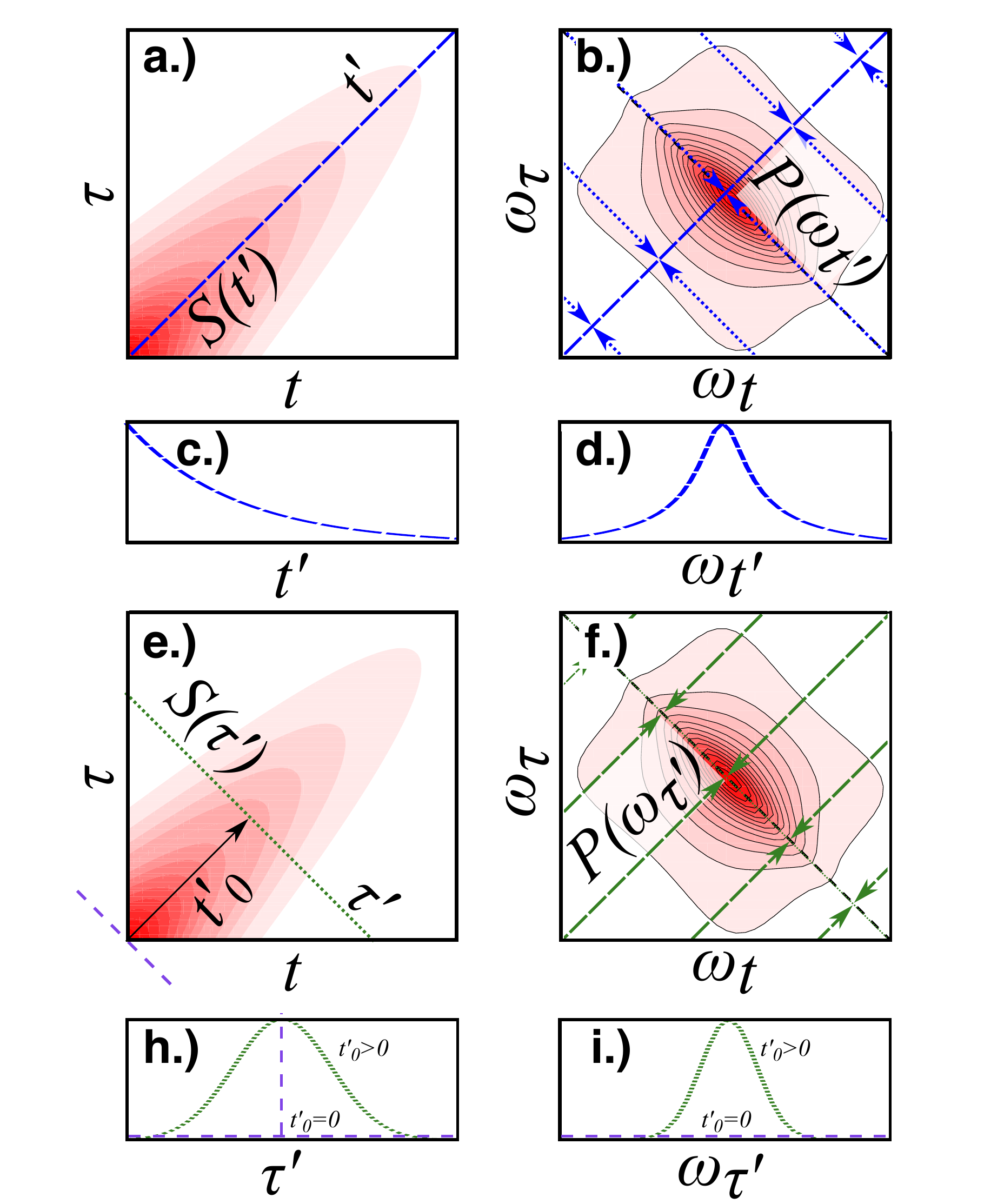} 
\caption{Diagram showing the rotated coordinate system in the MDCS time [a.), e.)] and frequency domains [b.), f.)], along with the slices [c.), h.)], and associated projections [d.), i.)], that are the focus of this work. The dashed lines in a.) and e.) represent the data slices in the time domain and arrow tipped lines  in b.) and f.) represent bins that are integrated in the projections. All plots show the simulated rephasing amplitude.}
\label{fig:ProjSliceImageAnnotated}
 \end{figure}

The two orthogonal time axes $t' = \frac{\left ( t + \tau \right )}{\sqrt[]{2}}$ and $\tau' = \frac{\left (t - \tau \right )}{\sqrt[]{2}}$ (shown in Fig. \ref{fig:ProjSliceImageAnnotated}) correspond to the diagonal and cross-diagonal directions in the MDCS time domain, and allow Eq. \ref{eq:Signalttau} to be rewritten in this new basis. The signal normalized to $s_{0,0}$ in this new basis is 

\begin{equation} \label{eq:Signaltprimetauprime}
s(t',\tau')= e^{ - \left ( \sqrt[]{2} \gamma  t' + \, i  \, \, \sqrt[]{2} \omega_{0} \tau' + \, \sigma^2 \tau'^2 \right )} \Theta(t' - \tau') \Theta(t' + \tau').
\end{equation}
This form of the signal simplifies the contribution of homogeneous and inhomogeneous broadening to the diagonal ($t'$) and cross-diagonal ($\tau'$) linewidths, at the cost of adding complexity to the step functions involved.


The Fourier transform of time domain \emph{slices} provides simplified expressions in both the time and frequency domains. 

A slice along $t'$, at $\tau'=0$, gives
\begin{equation} \label{eq:tprimeslice}
S(t',\tau'=0)=e^{ - \sqrt[]{2} \gamma  t' } \Theta^2(t')
\end{equation}
and a slice along $\tau'$, at a fixed $t'= t'_{0}$, results in the expression
\begin{equation} \label{eq:tauprimeslice}
s(t'=t_{0}',\tau')= e^{ - \left (\sqrt[]{2} \gamma  t_{0}' + i \, \, \sqrt[]{2}\omega_{0} \tau' + \sigma^2 \tau'^2 \right )} \Theta(t_{0}' - \tau') \Theta(t_{0}' + \tau'),
\end{equation}
where 
$t'_{0}$ is the intercept of the slice on the $t'$ axis. We note here that $t'_{0}$ \textit{must} be greater than zero to retrieve any 
meaningful information from a dataset that does not extend to negative delays. 
As shown by the purple dashed line in Fig. \ref{fig:ProjSliceImageAnnotated} [e.), h.), i.)], a slice along $t'_{0} = 0$ gives a delta function in time that when Fourier transformed becomes a constant in the frequency domain. Likewise, a slice at $t'_{0} < 0$ will be zero everywhere due to the step functions enforcing causality. 

Fourier transforming 
Eq. \ref{eq:tprimeslice} gives
\begin{equation} \label{eq:tprimeproj}
P(\omega_{t'})=\frac{1}{\sqrt[]{ 2 \pi} \left ( \sqrt[]{2} \gamma - i \omega_{t'} \right )},
\end{equation}
with an absorptive (dispersive) Lorentzian component for the real (imaginary) part of the lineshape. The absolute value of this 
complex lineshape is a square root Lorentzian, with a full-width at half-maximum (FWHM) of $\sqrt[]{2}\gamma$. The Fourier transform of the $\tau'$ slice gives 

\begin{equation} \label{eq:tauprimeproj}
\begin{aligned}
P(\omega_{\tau'}) &=e^{- \sqrt[]{2} t_{0}' \gamma} \, \, \frac{e^{- \left (\frac{\sqrt[]{2} \omega_{0} - \omega_{\tau'}}{2 \sigma} \right )^2}}{4 \sigma} \\ 
& \bigg ( Erf \left [ t_{0}' \sigma + \frac{i \left (\sqrt[]{2} \omega_{0} - \omega_{\tau'} \right )}{2 \sigma} \right ] 
\\
&+ Erf \left [ t_{0}' \sigma - \frac{i \left ( \sqrt[]{2} \omega_{0} - \omega_{\tau'} \right )}{2 \sigma} \right ] \bigg )\\
\end{aligned}
\end{equation}
which does have a $\gamma$ 
dependent term, but only as a constant scaling factor that does not affect the lineshape. Here $Erf$ denotes an error function. 
The expression in Eq. \ref{eq:tauprimeproj} has a 
Gaussian lineshape with a FWHM 
of $4 \, \, \sqrt[]{ln \left ( 2 \right )} \sigma$.


The expressions in Eqs. \ref{eq:tprimeproj} and \ref{eq:tauprimeproj} 
are analytical \emph{projections} onto the $\omega_{t'}$ and $\omega_{\tau'}$ axes, respectively. The use of the projection-slice theorem to arrive at these expressions is similar to the treatment of NMR spectra in \cite{Nagayama1978} with the important distinction that we use only the $t'$ and $\tau'$ directions to take advantage of their isolated $\gamma$ and $\sigma$ dependent behavior.

These expressions show the advantage of using the frequency projection as the basis for lineshape analysis: complete separation of the inhomogeneous and homogeneous broadening 
via their independent axes. 
These expressions thus improve upon previous analysis \cite{Siemens2010a,Bell2015} that resulted in \emph{coupled} expressions for the different broadening mechanisms.

The disadvantage of DS-FWM is that taking a projection 
removes the individual information content providing only an average material response \cite{Moody2011a}. 
Thus projections onto the frequency axis may be a more natural basis for considering single resonances or ensemble responses as a whole, while slices along the frequency axis are better suited to studying the response of individual oscillators in an ensemble.

\begin{figure}

   \centering
   \includegraphics[width=\linewidth]{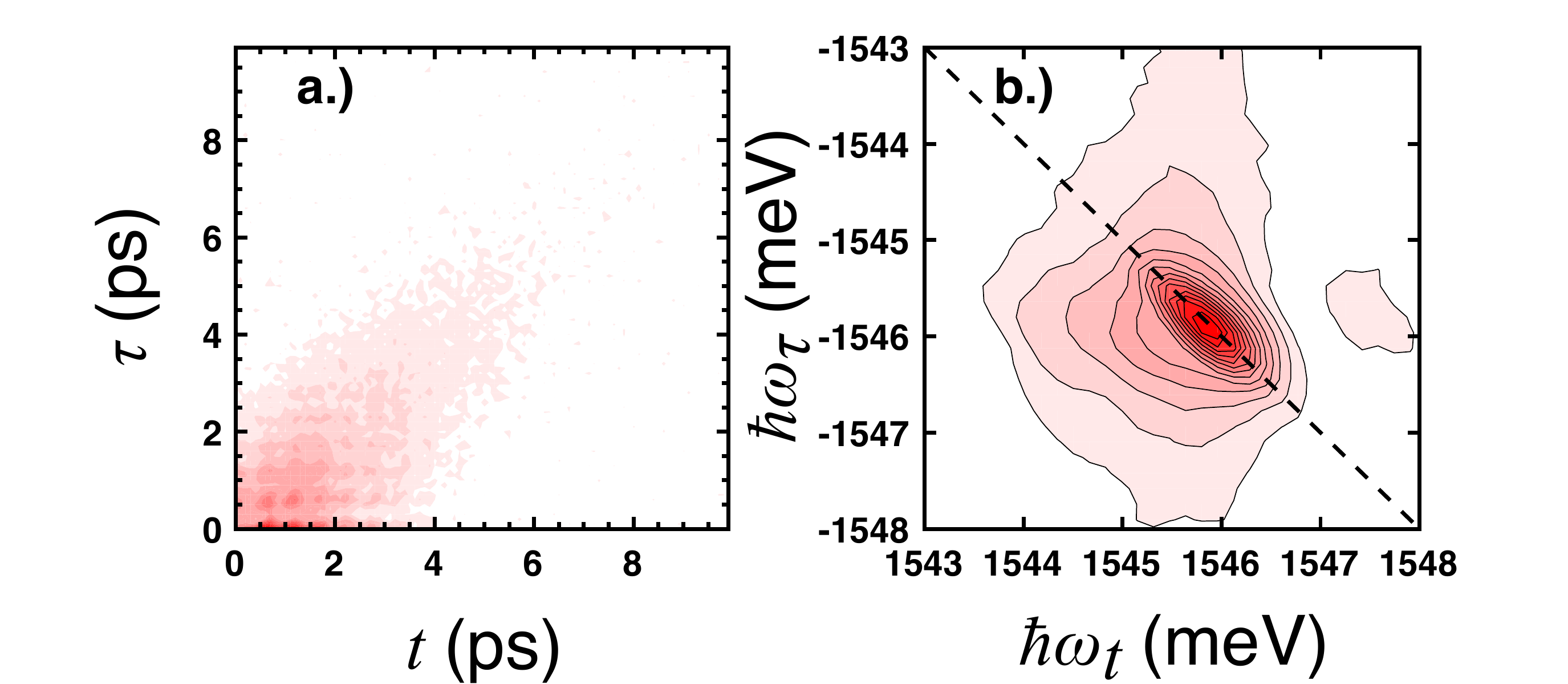} 
\caption{Experimental MDCS data from \textit{GaAs} QWs in the a.) Time and b.) frequency domain corresponding to the DS-FWM data shown in Figs. \ref{fig:DiagsliceData} and \ref{fig:CrsDiagsliceData}. Both spectra are absolute value rephasing spectra.}
\label{fig:2DData}
 \end{figure}


To 
validate the 
derived expressions 
we 
use them to fit simulated and experimental data. The experimental setup used is described in 
\cite{Nardin2013,Tekavec2007}, with the signal collected via photoluminescence (PL). 
This signal choice provides a direct analog to optical Ramsey spectroscopy in atomic physics. 
Briefly, a pulsed Ti:Sapphire oscillator (Spectra-Physics \textit{Tsunami}), with $ 90 \,fs$ pulses and a bandwidth of $ 30 \,meV$ is split into four identical copies, each with a precisely controlled delay via mechanical translation stages. Each beam is passed through an acousto-optic modulator (AOM) (Isomet \textit{1205-C}) where it is 
given a unique carrier frequency shift with a distinct radio frequency (RF). The resulting pulse train undergoes 
dynamic, pulse to pulse, phase cycling that averages out unwanted signal contributions and forces the signal PL to beat at RF frequencies specific to the desired quantum pathway.

This signal choice requires that all data must be collected in the time domain, point by point. This can significantly increase the number of points, and 
hence the acquisition time to acquire a MDCS
spectrum as compared to measurements using a spectrometer. However, by only collecting data along the $t'$ and $\tau'$ directions, we can 
measure the homogeneous and inhomogeneous linewidths in a similar amount of time that a coherently detected MDCS experiment could with no ambiguity or mixing of the broadening contributions. DS-FWM provides a greatly simplified analysis to extract many of the most important physical parameters accessible with MDCS. We realize data collection along the $t'$ and $\tau'$ axes by 
simultaneously stepping the stages that control $t$ and $\tau$ time delays in our experiment and collecting data at each point. This treatment is similar to the radial sampling used in some NMR experiments to reduce data collection time \cite{Kupce2005}, with the difference that we do not reconstruct full MDCS datasets from our projections.

 \begin{figure}
   \centering
   \includegraphics[width=\linewidth]{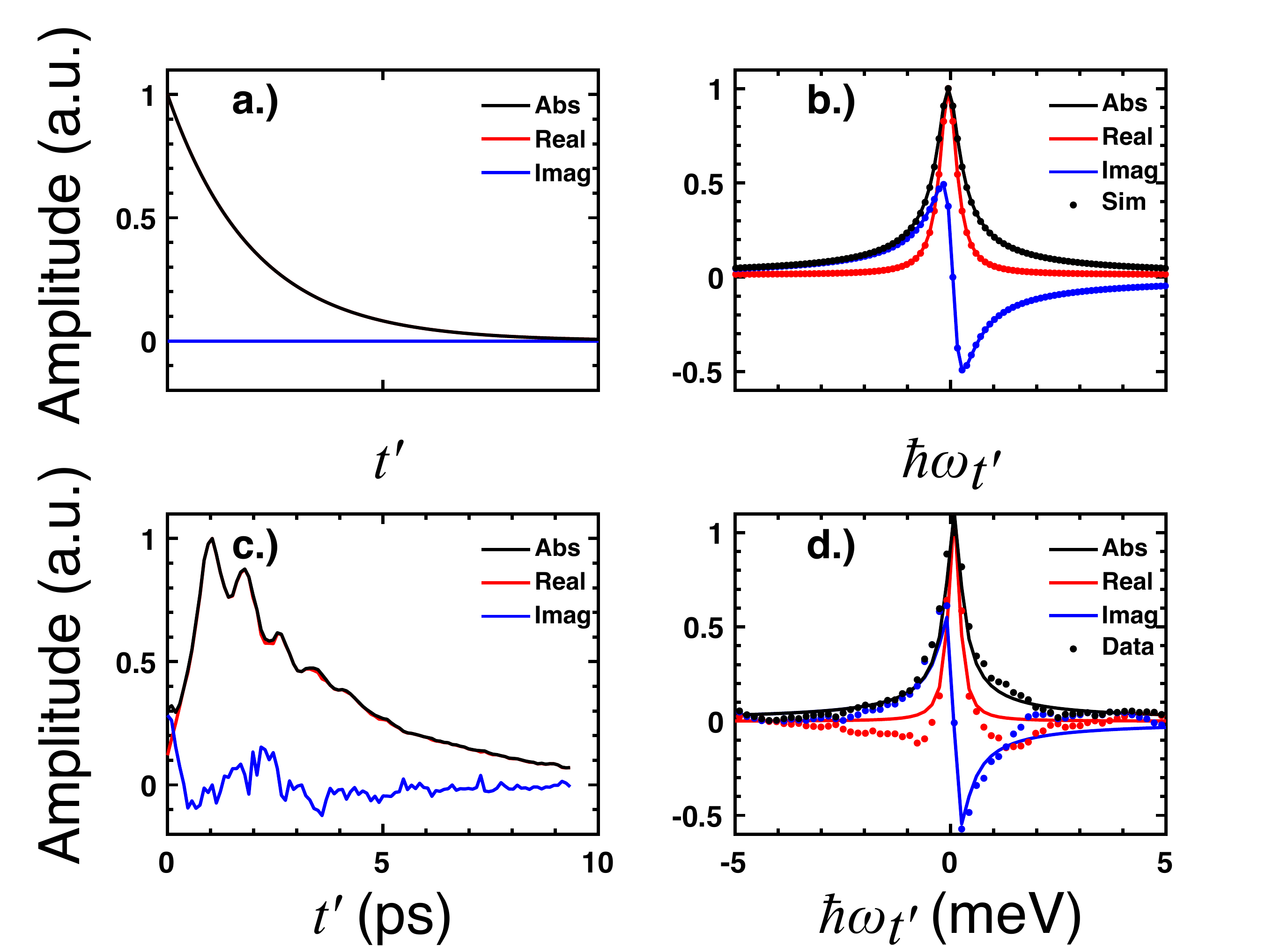} 
\caption{Time [a.), c.)] and frequency [b.), d.)] domain DS-FWM data for a simulated resonance [a.), b.)] and \textit{GaAs} QWs [c.), d.)]. All data was taken in the $\tau'$ direction and fit using Eq. \ref{eq:tprimeproj}. Dots show the data and solid lines show the function with best fit parameters $\left ( \gamma = 0.1051 \, meV \right)$.}\label{fig:DiagsliceData}
 \end{figure}

\begin{figure} 

   \centering
   \includegraphics[width=\linewidth]{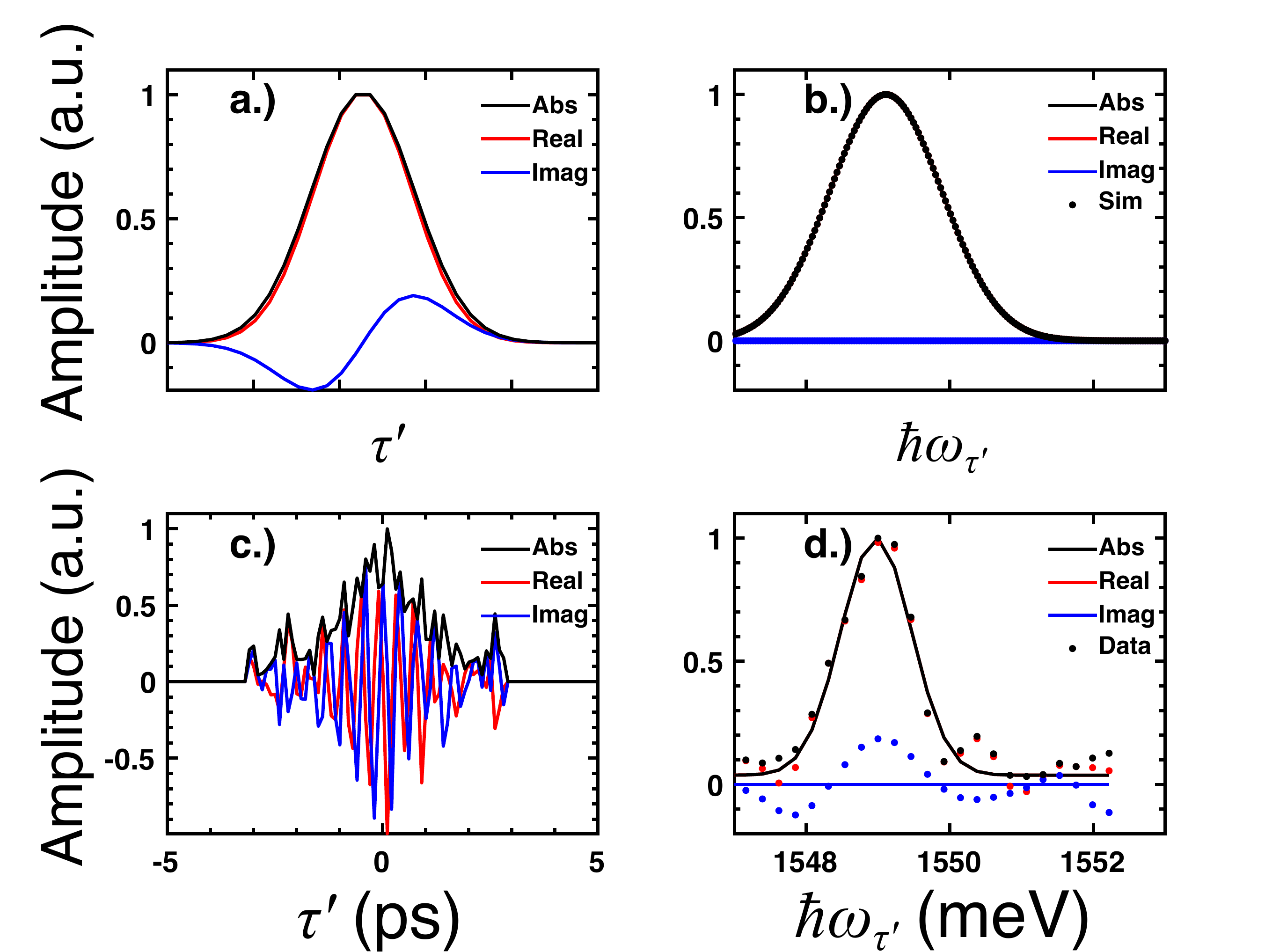}
\caption{Time [a.), c.)] and frequency [b.), d.)] domain DS-FWM data for a simulated resonance [a.), b.)] and \textit{GaAs} QWs [c.), d.)]. All data was taken in the $\tau'$ direction and fit using Eq. \ref{eq:tprimeproj}. Dots show the data and solid lines show the function with best fit parameters $\left ( \omega_{0} = 1548.97, \, \sigma = 0.306 \, meV \right )$.}\label{fig:CrsDiagsliceData}
 \end{figure}

To verify the derived 
DS-FWM expressions, we fit both simulated and experimental data to our complex functions. The sample used in this experiment 
consists of four $GaAs$ QWs surrounded by $Al_{0.3}Ga_{0.7}As$ barriers. It has been previously shown that strained bulk $GaAs$ \cite{Wilmer2016} and 'natural quantum dots' \cite{Moody2011} can be present in QW samples, in addition to the light hole (LH) exciton that is present in the sample. Here, we isolate the resonance of the heavy-hole (HH) exciton in the well by checking its frequency in a PL spectrum (Ocean Optics \textit{USB2000}) and tuning the excitation frequency to the HH PL frequency, with as little overlap with the LH exciton as possible. The sample is kept at a temperature below 10K by a recirculating liquid Helium cryostat (Montana Instruments \textit{Cryostation}). For simulated data of a single resonance with $\frac{\sigma}{\gamma} \approx 3$, both the diagonal, and cross-diagonal, projection is fit with $r^2 \approx .9999$. We then took DS-FWM spectra, in both the $t'$ and $\tau'$ directions. 
The experimental $\omega_{t'}$ projection was fit to our expressions with $r^2 \approx .9682$ and the $\omega_{\tau'}$ projections fit the data with $r^2 \approx 0.9718$. The results of these fits can be seen in Figs. \ref{fig:DiagsliceData} and \ref{fig:CrsDiagsliceData}. We acknowledge that the sample system was not the ideal single resonance 
assumed by the derivation, as can clearly be seen in the beating of the \textit{GaAs} QW time-domain data in Figs. \ref{fig:2DData} and \ref{fig:DiagsliceData}. The LH exciton of the quantum well has a small signal that is present in the DS-FWM spectra but is much weaker than the HH resonance and should not significantly contribute to the projected lineshapes. Furthermore, we note that we have normalized the amplitude of our projections to unity in our treatment of the data. Normalizing the data in this way forfeits our ability to extract oscillator strengths from our fits, as we are focused on the broadening contributions of the resonance.

We note that this procedure may be performed directly by the experiment in the time domain by only collecting data in the $t'$ and $\tau'$ directions or in the analysis of a full MDCS spectrum by extracting data slices from a larger data set. We have performed our analysis using each of these procedures and seen no significant difference in the results. 
We also note that any MDCS experiment that collects all data in the time domain, such as those that detect a population with a fourth readout pulse as opposed to an emitted electric field or those that use a fourth pulse for heterodyne detection on a photodiode,
is already set up to take DS-FWM spectra, 
although the potential of 
a simplified experimental setup that collects data slices only along $t'$ and $\tau'$ in the MDCS time domain is very promising. Such an experiment could access critically relevant material parameters $\left ( \gamma, \sigma, \omega_{0} \right )$ without the data collection times and experimental complexity of many MDCS experiments.

In this letter, we have presented a data collection scheme for ultrafast coherent spectroscopy and an associated lineshape analysis that is applicable to MDCS spectra as well as DS-FWM spectra. We have derived analytical expressions for slices along the $t'$ and $\tau'$ axes in the MDCS time domain, as well as expressions for the associated frequency domain $\omega_{t'}$ and $\omega_{\tau'}$ projections and shown that these projections completely separate the homogeneous and inhomogeneous broadening mechanisms to the lineshape. We have fit these expressions to both simulated and experimental data and shown excellent agreement. The technique presented here offers a deeper insight into the nature of lineshape broadening in coherent spectroscopy as well as a protocol for faster data collection to find key material parameters.




National Science Foundation (NSF) (1511199, 1553905).
 
The authors thank Christopher Smallwood and Matthew Day for advice on photoluminescence detection and for sharing their design of a custom amplifier circuit, as well as Samuel Alperin and Jasmine Knudsen for helpful discussions concerning the manuscript. T.M.A. acknowledges support from a National Research Council (NRC) Research Associate Program (RAP) award at the National Institute of Standards and Technology (NIST).

©  2018 Optical Society of America. Users may use, reuse, and build upon the article, or use the article for text or data mining, so long as such uses are for non-commercial purposes and appropriate attribution is maintained. All other rights are reserved.

\bibliography{Paperbibliographies-Projection_slice_paper_references.bib}
\bibliographystyle{unsrt}


\end{document}